# SocialStegDisc: Application of steganography in social networks to create a file system

Jędrzej Bieniasz
Institute of Telecommunications
Warsaw University of Technology
Warsaw, Poland
J.Bieniasz@stud.elka.pw.edu.pl

Krzysztof Szczypiorski
Institute of Telecommunications
Warsaw University of Technology
Warsaw, Poland
ksz@tele.pw.edu.pl

*Abstract—* The concept named SocialStegDisc was introduced as an application of the original idea of StegHash method. This new kind of mass-storage was characterized by unlimited space. The design also attempted to improve the operation of StegHash by trade-off between memory requirements and computation time. Applying the mechanism of linked list provided the set of operations on files: creation, reading, deletion and modification. Features, limitations and opportunities were discussed.

## I. INTRODUCTION

In an increasing number of scenarios of digital economy, security mechanisms in storage, modification or sharing of sensitive data require to be considered and applied. Most IT systems use external mass storage devices (not embedded directly in the processor) to store data of various sizes. Access to and management of their physical form is ensured by a filesystem.

An external storage device, referred to as a drive, is deemed to be a safe drive when physical and software implementation of a given drive performs security services: confidentiality, authenticity and integrity of data. The fourth desired feature of safe drives is hiding performance of basic operations in a filesystem. Steganography may be a way of hiding communication with drives or performance of drive operations. The purpose of steganography is to hide information in a carrier of this information in a way which is unnoticeable to the observer.

Throughout the years, application of steganography in mass storage has evolved into a separate field of research referred to as File System Steganography. A milestone in this research area was the work of Anderson, Needham and Shamir [1]. They proposed a steganographic filesystem allowing to deny the existence of entire file directories. In general, steganographic filesystems are an application of steganography, therefore their development is often preceded by appearance of new steganographic techniques. This work is an example of such coupling. In 2016 a new type of a steganographic technique called StegHash was proposed in [2]. In this system, multimedia files placed in open social networks (OSNs) are a carrier of hidden information. The essence of the system is the existence of a logical connection between them by means of a mechanism of hashtags, that is text markers in the form of #<tag>, commonly applied in OSNs. In addition, the author of [2] proposes an option of applying the StegHash technique to create a steganographic filesystem analogous to the existing classic filesystems such as FAT (File Allocation Table) or NTFS (New Technology File System). This work focuses on extending this application of the StegHash method by proposing a new type of technique called SocialStegDisc.

## II. STATE OF THE ART

### A. Steganographic filesystems

As mentioned above, a breakthrough in the development of steganographic filesystems is an article written by Anderson, Needham and Shamir [1]. The authors proposed a method utilizing the way in which invisible ink works. A user with a password is able to find and decode hidden information. The StegFS system presented in [3] is a practical implementation of the concept presented in [1] as an extension of a standard filesystem, Linux Ext2fs.

In recent years, distributed filesystems used in clusters of various kinds have been gaining popularity. For such systems, proposals of applying steganography to hide information in them also appear, for instance in [4]. The main idea behind the method presented in [4] consists in using multiple files to encode hidden information by means of their relative positions in clusters.

Among recent proposals, work presented in [5] is worth highlighting, in which Neuner et al. presented a way of hiding information in timestamps of files saved in classic filesystems. They discovered that information may be effectively hidden in 24 bits (3B) coding the nanosecond part of the timestamp in the NTFS filesystem. For each file, it is possible to use two types of timestamps: of creation and of last access. In this way, one file creates 6B of space for hidden data.

### B. Use of OSNs in steganography

As soon as open social networks appeared, steganography researchers began searching for options to apply them in steganographic systems.

In [6] the authors define two types of communication: with high and low entropy. In a high entropy model, multimedia files such as images, movies or music are data carriers. This is a classic example of multimedia steganography, where a single object carries hidden information. A covert channel of this type is characterized by high throughput but is easy to detect. A low entropy model consists in using text data to carry hidden information. In addition, information from such a hidden information carrier is decoded by means of a previously known method (secret). A hidden channel of this type is characterized by a low throughput but a good level of security. The authors also propose using such a channel to establish the location of the appropriate steganogram, and they indicate Internet services, for instance social networks, as a carrier. Proposals of low entropy steganographic methods are presented in [7]. The first one uses file names as a hidden information carrier and requires OSNs not to change them. For this method, the authors propose using standard file naming schemes by photographic devices. Sequence numbers in these names are a hidden data carrier. The second method assumes using tags in pictures. Numerous images with numerous users' tags creates a hidden channel. On the basis of a prepared set of pictures and a set of tags, it is possible to develop a matrix encoding hidden information. Other approaches based on language features are presented in [8] and [9]. Steganograms are transferred as bitmaps coded by means of language permutation. The created channel is very safe but there is a high likelihood of its distortion.

All these methods use a classic multimedia steganography, easy to detect, or a more sophisticated one, with a limited steganographic throughput. The proposed methods assume using multiple user accounts in OSNs and a parallel access to an abundance of files, which may be an insecurity indicator on the servers' side and an operation blockage.

### III. CONCEPT OF THE SYSTEM

There is a basic limitation to applying StegHash in SocialStegDisc: memory occupancy. Along with the increase of the number of $n$ hashtags, the volume of the dictionary increases proportionately to $n!$. Therefore, keeping the entire dictionary in the memory places a heavy burden on the system, and generation of the dictionary and restoring the set of used addresses will consume a large amount of time. The main goal behind keeping these dictionaries is ensuring uniqueness and reliability of addressing. Therefore, for a higher $n$ it is proposed to introduce a dynamic verification of uniqueness of the generated addresses. Two space-time tradeoff-type modifications will be formulated on its basis along with an application of the linked list mechanism.

#### A. Dynamic verification of uniqueness of addresses

The first attempt to increase the efficiency of the StegHash method is to eliminate storing results from generation of the utilized address space when creating a chain and later restoring it. Such a set was stored because it was necessary to verify the uniqueness of the generated addresses. In this proposal, uniqueness is verified by checking in the OSNs whether there is a file with a given address.

#### B. Space-time tradeoff modifications and linked list mechanism

From the perspective of a correct system operation, the system of addressing, that is indexing SocialStegDisc segments with hashtags, is of key importance. A characteristic feature of such addressing is non-linearity. Subsequent addresses are generated by means of a specific function on the basis of the initial address. The issue of addressing in SocialStegDisc is an analogy to the problem of memory space fragmentation in classic filesystems. Until now, a strong assumption was a prohibition of hiding addresses of the system due to security reasons. This implies that it is not possible to apply mechanisms servicing memory fragmentation, for instance allocation tables or storing the index to another memory block together with the data. As a consequence, it is not possible to make available a function of deleting single files because an occurrence of any gap in addressing, if it is not known that the gap exists, would mean that the files located after the gap would be lost. Under this approach, once a file is added to SocialStegDisc, it will remain there. In this article, a solution is proposed which would allow to introduce structures for efficient management of resources (addresses), known from classic filesystems, to SocialStegDisc. To this end, the assumption of non-storing the addresses will be weakened by introducing such a mechanism which will simultaneously not decrease the security level.

Direct addresses to objects may be presented by means of an address to the object of reference (beginning of the SocialStegDisc instance) and a counter, whose value will serve to control obtaining the address of the subsequent object. In every multimedia object, space for this counter is added to hidden space. Due to this, the SocialStegDisc system assumes the structure of a linked list which allows to supplement the system with file deletion. An example of how this operates is presented in the figure below.

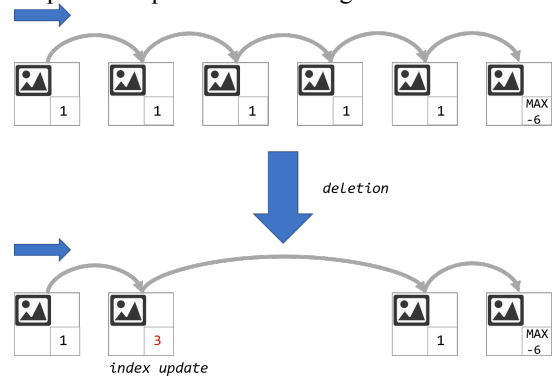

Figure 1. Mechanism of a linked list in SocialStegDisc

Apart from hiding $m$ bytes of proper data, it is necessary to place a minimum of $p$ bits coding the numerical index in every multimedia object. The selection of the volume $p$ determines the maximum length of the chain, for which the problem of information loss does not occur when the mechanism of file deletion is applied, and amounts to $2^p - 1$. The hidden counter will be used in proposed modifications of the space-time tradeoff type, by using it to control the functions establishing the next address. This counter may store:
1) the address code for application of the StegHash method without the dictionary of used addresses;
2) the counter value which specifies how many times the procedure of generation of a subsequent address should be performed to actually obtain it – in the case of elimination of both original dictionaries of the StegHash method.

When StegHash with a modification eliminating the dictionary of used addresses is applied, when a chain of objects is created, a code of the address to the next object is placed in the additional space, in line with the coding from the dictionary of permutations. The scheme of this procedure is presented in Figure 2.

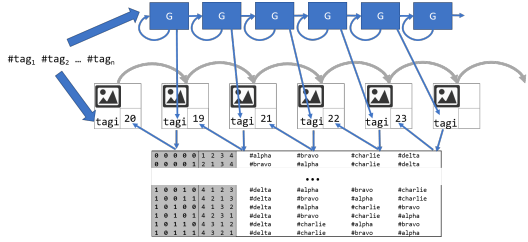

Figure 2. Creation of a chain by means of combined StegHash method and linked list mechanism

During reading, the address code is retrieved from the hidden space and then converted into an address in the form of a set of hashtags. Thus, access to the next object is possible. To maintain the continuity of the address space, when deleting a file in the multimedia object preceding the beginning of the given file, the counter should be updated to the value retrieved from the last object storing the file to be deleted. The set of the used addresses should also be updated. The released addresses are not lost but they go back to the pool and may be generated in the subsequent iterations of pseudo-random generation function.

Another proposed modification of space-time tradeoff type is a total elimination of original StegHash dictionaries and using the linked list mechanism to store relevant counters allowing to restore the chain. To this end, first of all, the algorithm of dynamic generation of addresses should be modified so that the function, apart from the new address, returns the current status of the sampling counter in the course of generation of subsequent addresses. With the starting point known, the obtained counter allows to retrieve the address returned in the course of generation at the counter status. As these states of the counter are recorded in the hidden space defined for the linked list mechanism, they allow to traverse the chain objects from a known initial address of the SocialStegDisc instance. In Figure 3, a scheme of a chain creation is presented.

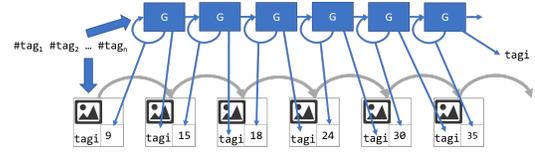

Figure 3. Chain creation with no dictionaries and a linked list mechanism

File deletion which entails creation of free address space, in this case updates the index in the object preceding the space to the value of the index from the last deleted file.

IV. DISCUSSION

The issue of addressing in SocialStegDisc is an analogy to the problem of memory space fragmentation in classic filesystems, which has been highlighted when developing the concept. The SocialStegDisc concept introduces a mechanism of storing indexes to the next object in the system, due to which the drive assumes the structure of a linked list. Thanks to the application of a linked list mechanism, it is possible to efficiently use the space and propose a file deletion mechanism. The greatest problem is the calculation overhead and algorithms verifying the correctness of the operation scheme.

The system is characterized by non-linearity in filling in the byte space. This effect occurs when the last used object hides a smaller number of bytes than possible. On the basis of the number of bytes, it is possible to determine the length of the chain of multimedia files created in line with the StegHash technique essence. This size amounts to:

$$L = \left\lceil \frac{M}{m} \right\rceil \quad (1)$$

where $M$ is the number of bytes on which the operation is performer, and $m$ is the size of a single logical block. When $M \bmod m \not\equiv 0$ in the last multimedia file in the chain, fewer bytes are placed than the size of space allocated in the multimedia file ($m$).

The SocialStegDisc undetectability level depends on:
1) the level of undetectability by open social networks of applied methods of multimedia steganography, serving to hide data;
2) the level of undetectability of the StegHash steganographic method which is a system of indexing multimedia objects and building the SocialStegDisc drive space from single blocks;
3) the level of security of social networks with respect to detection of anomalies in network users' behaviour. The SocialStegDisc system may communicate with the same service multiple times at short intervals. From the perspective of such a service, frequent and automated sending of queries will be treated as a threat, for instance as an attack of a Denial-of-Service type;
4) the level of detectability of multimedia steganography. This problem was under research for Facebook, among others. In [10], the authors tested algorithms sanitizing steganography on the side of Facebook servers. The research shows that Facebook algorithms are efficient in the detection of steganography known so far, in particular for hidden information of greater size.

In [11], a way of evaluation of network steganography techniques was proposed in three categories of undetectability: good, bad and ugly. The presented classification manner may be applied to other steganographic methods. StegHash may be considered a good steganographic method, which transitively means that the proposed SocialStegDisc technique is also characterized by this feature. The observer is unable to detect the ongoing communication anywhere in the network, even at the receiver of hidden data.

## V. IMPLEMENTATION

### A. Environment

The system implementation proposal presented in this article is a proof-of-concept of the SocialStegDisc system. In Figure 4 a scheme of operation of the developed testing environment is presented.

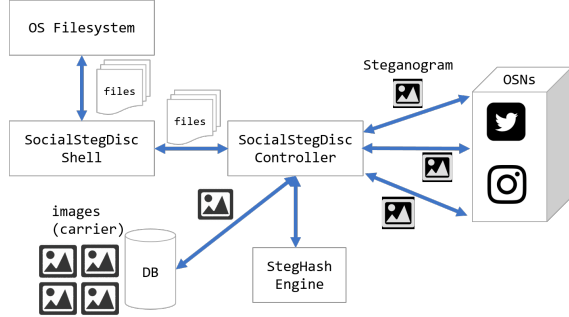

Figure 4. Scheme of operation of SocialStegDisc

The main client application is the SocialStegDisc Shell terminal, which translates the client queries to appropriate operations of the file directory and operations on files, using the Controller module. In addition, two modules known from StegHash [2] have been used:
1) StegHash Engine – a module responsible for the creation of appropriate addressing of multimedia files with the use of hashtags.
2) Database – contains a set of multimedia files used as a carrier of hidden data.

### B. Generation of addresses

The basic operation of the system is generation of addresses. This index plays two roles:
1) Unambiguously identifies the object,
2) Identifies the social network in which it is embedded.

In the StegHash system [2], addresses were generated by means of any type of permutation generation algorithm, and uniqueness was ensured by rejecting the already drawn addresses. For SocialStegDisc, generation of addresses in a chain must be characterized by restorability with the use of appropriate initial conditions. To this end, an approach of Rivest [12] was used, who presented a reference code for a program generating a random sample of $s$ size from set $S$. Assuming that $s = S$, the sample is a searched permutation of set $S$. In Figure 5, an exemplary code in Python of a chain generation method with no dictionaries stored is presented.

```
def go_to_perm(seed, counter):
    input = ""
    for num in seed:
        input += str(num)
    i = 0
    output_list = []
    while(i < counter):
        if(len(output_list) == len(seed)):
            for num in output_list:
                input = ""
                input = str(num)
                output_list = []
        i = i + 1
        hash_input = input + str(i)
        hash = hashlib.sha256(hash_input.encode('utf-8'))
        output = int(hash.hexdigest(),16)
        pick = int(output % len(seed))
        if not pick in output_list:
            output_list.append(pick)
    return output_list
```

Figure 5. Method of address generation based on reference address

The main part of the method is using the hash function for pseudorandom generation of the next value which is a candidate to become an element of an output permutation. In the example of the code, the function SHA-256 was used [13], but it is possible to apply other hash functions, for instance SHA-512 [13]. Using a hash function with a greater output size reduces the number of sampling iterations when completing the output table. The output value from the hash function is converted to a hexadecimal number, based on which a decimal value is generated, which is a pseudorandom number generated in a given loop iteration. This number undergoes the operation of obtaining remainder from division by the set size. At the end, it is verified whether the sampled value already exists in the output set from a given iteration of permutation generation. If not, it should be added to the list. Generation ends when the table of numbers has been obtained of the size of a set undergoing permutation. The returned table of numbers is mapped to relevant hashtags.

### C. Hosting in social media

When selecting a way of multimedia hosting, attention should be paid to multimedia files processing on the side of social network services and the impact this processing has on non-breach of hidden data. This problem was researched for, among others, Facebook. In [10], the authors verified various approaches to eliminate the compression effect and algorithms sanitizing steganography on the side of Facebook servers. Research shows that Facebook algorithms are efficient in detecting the steganography known so far, in particular for hidden information of greater size. Therefore, it should be verified first whether the social networks selected for the needs of practical implementation of SocialStegDisc ensure non-interference with the byte content of files.

Problems occurring with sanitization of steganography on the side of social networks or with its deformation may be circumvented by using an intermediary layer in the form of file upload services. An advantage of such approach is that such service has no impact on the file content, and thus – there is a certainty of non-breach of steganographic content. Then, in the social service using a hashtag mechanism only a link and a set of hashtags addressing this file are shared.

## VI. Summary

In this work, the authors' own concept of a new steganographic filesystem called SocialStegDisc was formulated. This system works in analogy to classic filesystems such as FAT or NTFS. A carrier of hidden information are multimedia files embedded in various open social networks. Every multimedia object is indexed with a unique permutation of a set of *n* hashtags. Between the objects there is a logical structure created by a set hashing function. On the basis of an input index (permutation of a set of hashtags), this function generates an index of the next object (permutation of a set of hashtags). Thus, subsequent blocks of hidden space may be read, recorded or modified in sequences, which is an analogy to servicing classic filesystems.

Researching such issues as the SocialStegDisc technique focuses on their usefulness in real scenarios, especially with respect to information leak. On one hand, it may seem that development of such concepts may serve only terrorists or other criminals in their attempts to hide communication supporting their actions. But above all, the purpose of research on this method was to demonstrate its application options and to determine on this basis the vulnerability of the systems used in it. Such vulnerability may be taken into account at the following iterations of the system development and be entirely eliminated.